%% file: pub_0110_la.tex
\documentclass[twocolumn,showpacs,aps,prl,superscriptaddress]{revtex4b5}

\usepackage{graphicx}
\usepackage{dcolumn}
\usepackage{amsmath}
\usepackage{epsfig}

\input{pubboard/babarsym}
\input{symbols}

\newcommand{\BABARPubYear}    {01}
\newcommand{\BABARPubNumber}  {10}
\newcommand{\SLACPubNumber} {8838}

\def\figurebox#1#2#3{%
    \def\arg{#3}%
    \ifx\arg\empty
    {\hfill\vbox{\hsize#2\hrule\hbox to #2{\vrule\hfill\vbox to #1{\hsize#2\vfill}\vrule}\hrule}\hfill}%
    \else
    {\hfill\epsfbox{#3}\hfill}%
    \fi}

\long\def\inst#1{\par\nobreak\kern 4pt\nobreak
    {\it #1}\par\vskip 10pt plus 3pt minus 3pt}

\begin{document}

\preprint{\babar-PUB-\BABARPubYear/\BABARPubNumber}
\preprint{SLAC-PUB-\SLACPubNumber}

\begin{flushleft}
\babar-PUB-\BABARPubYear/\BABARPubNumber\\
SLAC-PUB-\SLACPubNumber\\
\end{flushleft}

\title{
\vskip 10mm
{
\Large \bf
Measurement of Branching Fractions and Search for 
{\boldmath \CP}-Violating Charge Asymmetries in
Charmless Two-Body {\boldmath \B} Decays into Pions and Kaons
}
\begin{center} 
\vskip 10mm
The \babar\ Collaboration
\end{center}
}

\input{authors_0110}

\date{May 21, 2001}

\begin{abstract}
We present measurements of the branching fractions and a
search for \CP-violating charge
asymmetries in charmless hadronic decays of $B$ mesons into two-body 
final states of kaons and pions.  The results are based on a data
sample of approximately $23$ million \BB\ pairs collected by 
the \babar\ detector at the
\pep2\ asymmetric $B$ Factory at SLAC.  We find the following branching
fractions:
$\BR(\btopp)= (4.1\pm 1.0\pm 0.7)\times 10^{-6}$,
$\BR(\btoKpp)=(16.7\pm 1.6\pm 1.3)\times 10^{-6}$,
$\BR(\btoKpz)=(10.8^{+2.1}_{-1.9}\pm 1.0)\times 10^{-6}$,
$\BR(\Bu\to \Kz\pip)=(18.2^{+3.3}_{-3.0}\pm 2.0)\times 10^{-6}$,
$\BR(\btoKzpz)=(8.2^{+3.1}_{-2.7}\pm 1.2)\times 10^{-6}$.
We also report the $90\%$ confidence level
upper limits $\BR(\btoKK)< 2.5\times 10^{-6}$,
$\BR(\btoppz) < 9.6\times 10^{-6}$, and
$\BR(\btoKzK)<2.4\times 10^{-6}$.  In addition, charge asymmetries
have been measured and found to be
consistent with zero, where the statistical precision is in the range
of $\pm 0.10$ to $\pm 0.18$, depending on the decay mode.
\end{abstract}

\pacs{
13.25.Hw, 
13.25.-k, 
14.40.Nd  
}

\maketitle

\par
The study of $B$ meson decays into charmless hadronic final states plays
an important role in the understanding of \CP\ violation.
In the Standard Model, all \CP-violating phenomena
are a consequence of a single complex phase in the Cabibbo-Kobayashi-Maskawa
(CKM) quark-mixing matrix \cite{ckm}.
Recently, the Belle and \babar\ collaborations published 
results~\cite{bellestb,babarstb} on measurements of \CP-violating asymmetries 
in $B$ decays into final states containing charmonium, leading to constraints 
on the angle $\beta$ of the CKM Unitarity Triangle.  Measurements of the 
rates and charge asymmetries for $B$ decays into the charmless final states 
$\pi\pi$ and $K\pi$ can be used to constrain the angles $\alpha$ and 
$\gamma$~\cite{gamalpha} of the Unitarity Triangle.

In this Letter we present new measurements of the branching fractions for
$B$ meson decays to the charmless hadronic final states $\pip\pim$,
$\Kp\pim$, $\Kp\piz$, $\Kz\pip$ and $\Kz\piz$ \cite{cc}.
In addition, we search for charge asymmetries in the 
modes $\Bz\to \Kp\pim$, $\Bu\to \Kp\piz$ and 
$\Bu\to \Kz\pip$.  Previous
measurements \cite{cleobr,cleocp} of these decays were reported
by the CLEO Collaboration.
 
The data sample used in these analyses was collected with the \babar\ detector
\cite{babarnim} at the \pep2\ \epem\ collider \cite{pepii} at the 
Stanford Linear Accelerator Center.
It corresponds to an integrated luminosity of $20.6\invfb$ taken on the
\FourS\ resonance (``on-resonance'') and $2.61\invfb$ taken at a 
center-of-mass (CM) energy $40\mev$ below the \FourS\ resonance
(``off-resonance''), which are used for continuum background studies.
The on-resonance sample corresponds to $(22.57\pm 0.36)\times 10^6$ 
\BB\ pairs.  The
collider is operated with asymmetric beam energies, producing a boost
($\beta\gamma = 0.56$) of the \FourS\ along the collision axis ($z$).  
The boost increases the momentum range of two-body $B$ decay products 
from a narrow distribution centered near $2.6\gevc$ to a broad 
distribution extending from $1.7$ to $4.3\gevc$.

The \babar\ detector is a spectrometer of charged and neutral particles
and is described in detail in Ref.~\cite{babarnim}.  Charged particle (track)
momenta are measured in a tracking system consisting of a $5$-layer,
double-sided, silicon vertex detector and a $40$-layer drift
chamber (DCH) filled with a gas mixture of helium ($80\%$) and isobutane
($20\%$), both operating within a $1.5\,{\rm T}$ superconducting solenoidal
magnet.  Photons are detected in an electromagnetic 
calorimeter (EMC) consisting
of $6580$ CsI(Tl) crystals.  Charged hadron identification
is based on the Cherenkov angle $\theta_c$
measured by a unique, internally reflecting Cherenkov ring imaging
detector (DIRC). 

Hadronic events are selected based on track multiplicity and event topology.
Backgrounds from non-hadronic events are reduced by requiring the
ratio of Fox-Wolfram moments $H_2/H_0$ \cite{fox} to be less than $0.95$
and the sphericity \cite{spheric} of the event to be greater than
$0.01$.

All tracks (except \KS\ decay products) are required to have a polar angle
within the tracking 
fiducial region $0.41 < \theta < 2.54\,{\rm rad}$ and a Cherenkov
measurement from the DIRC.  The latter is satisfied by
$91\%$ of the tracks in the fiducial region.  We require a
minimum number of Cherenkov photons associated with each $\theta_c$
measurement in order to improve the resolution.  The efficiency of
this requirement is $97\%$ per track.  Tracks with a $\theta_c$ within
$3\sigma$ of the expected value for a proton are rejected.
Electrons are rejected based on specific ionization
($dE/dx$) in the DCH system, shower shape in the EMC, and the ratio of
shower energy to track momentum.

Candidate $\KS$ mesons are reconstructed from pairs of oppositely
charged tracks that form a well-measured vertex and have an invariant
mass within $3.5\sigma$ of the nominal \KS\
mass \cite{PDG}.  The measured proper decay time of the \KS\ candidate
is required to exceed $5$ times its error.

Candidate \piz\ mesons are formed from pairs of
photons with an invariant mass within $3\sigma$ of the nominal \piz\
mass.  Photons are defined
as showers in the EMC that have the expected lateral shape, are not
matched to a track, and have a minimum energy of $30\mev$.  
The \piz\ candidates are then kinematically fitted with their mass 
constrained to the nominal \piz\ mass. 

$B$ meson candidates are reconstructed in four topologies: $h^+ h^{\prime -}$,
$h^+\piz$, $\KS h^+$ and $\KS\piz$, where the symbols
$h$ and $h^{\prime}$ refer to $\pi$ or $K$.  
The kinematic constraints provided by
the \FourS\ initial state and relatively precise knowledge of the
beam energies are exploited to efficiently identify $B$ candidates.  
We define a beam-energy 
substituted mass $\mes = \sqrt{E^2_{\rm b}-\mathbf{p}_B^2}$, 
where $E_{\rm b} =(s/2 + \mathbf{p}_i
\cdot\mathbf{p}_B)/E_i$, $\sqrt{s}$ and
$E_i$ are the total energies of the \epem\ system in the CM
and lab frames,
respectively, and $\mathbf{p}_i$ and $\mathbf{p}_B$ are 
the momentum vectors in
the lab frame of the \epem\ system and the $B$ candidate, respectively.  
To improve the resolution in modes containing \piz\ mesons, the \B\ candidate
is kinematically fitted with the energy constrained to the CM beam energy. 
For all modes, the \mes\ resolution is dominated by the beam energy spread 
and is approximately $2.5\mevcc$.  
Candidates are selected in the range $5.2<\mes<5.3\gevcc$.

\begin{table*}[!htb]
\begin{center}
\caption{ Summary of results for detection efficiencies
($\varepsilon$), fitted signal yields ($N_S$), statistical 
significances ($S$), 
measured branching fractions (\BR), and
charge asymmetries.  
The efficiencies include the branching fractions for
$\Kz\to\KS\to\pip\pim$ and $\piz\to\gamma\gamma$.  Equal branching
fractions for \upsbzbz\ and $\Bu\Bub$ are assumed.  The $90\%$ 
confidence level (C.L.) intervals
for the charge asymmetries include the systematic uncertainties, which have
been added in quadrature with the statistical errors.}
\label{tab:brresults}
\begin{ruledtabular}
\begin{tabular}{lcccccccc} 
Mode~~~~  & ~~~$\varepsilon$ (\%)~~~ & ~~$N_S$~~ & ~~$S$ ($\sigma$)~~ & 
~~\BR($10^{-6}$)~~ &
~~~~\acp~~~~ & \acp\ $90\%$ C.L. \\ 
\hline
$\pip\pim$ &  $45$  & $41\pm 10\pm 7$ & 
$4.7$  & $4.1\pm 1.0\pm 0.7$
&   &  \\
$\Kp\pim$ &  $45$ & $169\pm 17\pm 13$ &
$15.8$ & $16.7\pm 1.6\pm 1.3$ 
& $-0.19\pm 0.10\pm 0.03$ & $[-0.35,-0.03]$ \\
$\Kp \Km$  & $43$ & $8.2^{+7.8}_{-6.4}\pm 3.5$  &
$1.3$  & $<2.5$ ($90\%$ C.L.) \\
\\
$\pip\piz$  & $32$ & $37\pm 14\pm 6$ & 
$3.4$  & $<9.6$ ($90\%$ C.L.) 
& &  \\
$\Kp\piz$    & $31$ & $75\pm 14\pm 7$ & 
$8.0$  & $10.8^{+2.1}_{-1.9}\pm 1.0$ 
& $0.00\pm 0.18\pm 0.04$ & $[-0.30,+0.30]$ \\
\\
$\Kz\pip$   & $14$ & $59^{+11}_{-10}\pm 6$ &
$9.8$  & 
$18.2^{+3.3}_{-3.0}\pm 2.0$ 
& $-0.21\pm 0.18\pm 0.03$ & $[-0.51,+0.09]$ \\
$\Kzb\Kp$    & $14$ & $-4.1^{+4.5}_{-3.8}\pm 2.3$  &
$-$    & $<2.4$ ($90\%$ C.L.) & & \\
\\
$\Kz\piz$  & $10$ & $17.9^{+6.8}_{-5.8}\pm 1.9$ &
$4.5$  & $8.2^{+3.1}_{-2.7}\pm 1.2$ 
& & \\
\end{tabular}
\end{ruledtabular}
\end{center}
\end{table*}

We define an additional kinematic parameter $\Delta E$ as the
difference between the energy of the $B$ candidate and half the energy
of the \epem\ system, computed in the CM system, where the pion mass
is assumed for all charged decay products of the \B.  
The $\Delta E$ distribution
is peaked near zero for modes with no charged kaons and shifted
on average $-45\mev$ ($-91\mev$) for modes with one (two) kaons, where
the exact separation depends on the laboratory kaon momentum.
The resolution on $\Delta E$ is mode dependent.  For final states that
contain no \piz\ mesons the resolution is about $26\mev$.  For modes
with \piz\ mesons the resolution is about $42\mev$ and is asymmetric
due to underestimation of the \piz\ energy in the EMC.  Candidates are
accepted in the following $\Delta E$ ranges (given in GeV): 
$[-0.15,0.15]$ ($h^+ h^{\prime -}$), $[-0.2,0.15]$
($h^+\piz$), $[-0.115,0.075]$ ($\KS h^+$) and $[-0.2,0.2]$
($\KS\piz$).

Detailed Monte Carlo simulation, off-resonance data, and events in
on-resonance \mes\ and $\Delta E$ sideband regions are used to study
backgrounds.  The contribution due to other $B$-meson decays, both
from $b\to c$ and charmless decays, is found to be negligible.  The
largest source of background is from random combinations of tracks and
neutrals produced in the $\epem\to \qqbar$ continuum (where $q=u$,
$d$, $s$ or $c$).  In the CM frame this background typically exhibits
a two-jet structure that can produce two high momentum, nearly
back-to-back particles, in contrast to the spherically symmetric
nature of the low momentum \upsbb\ events.

We exploit this topology difference by
making use of two event-shape quantities.  The first variable
is the angle $\thsph$~\cite{spheric}
between the sphericity axes of the $B$
candidate and of the remaining tracks 
and photons in the event.  The distribution
of $|\cos\thsph|$ in the CM frame is strongly peaked near $1$ for
continuum events and is approximately uniform for \BB\ events.  We
require $|\cos\thsph| < 0.9$, which rejects $66\%$ of the background
that remains at this stage of the analysis.

\begin{figure}[!tbh]
\begin{center}
\includegraphics[width=1.0\linewidth]{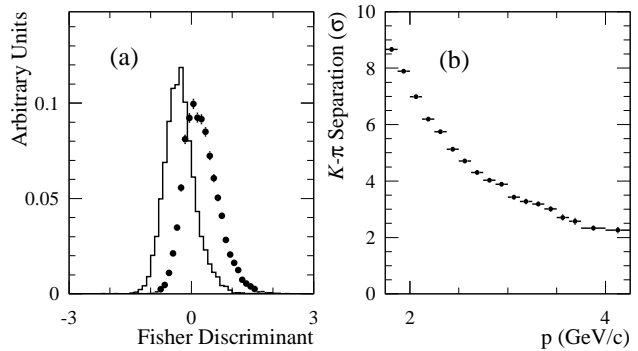}
\caption{
(a) The distributions of the Fisher discriminant for Monte Carlo simulated
$\Bz\to h^+ h^{\prime -}$ decays (histogram) and background events (points) in
the \mes\ sideband region $5.20<\mes <5.27\gevcc$;
(b) the $K$--$\pi$ separation, in units of standard deviations, as a
function of momentum, derived from the Cherenkov angle measurements 
of kaon and pion tracks in a $D^{*+}\to\Dz\pip$ control sample, as
described in the text.
}
\label{fig:fisher}
\end{center}
\end{figure}

The second quantity is a Fisher discriminant
${\cal F}$ constructed from the scalar sum of the CM momenta of
all tracks and photons (excluding the $B$ candidate decay products) flowing
into nine concentric cones centered on the thrust axis of the $B$
candidate.  Each cone subtends an angle of $10^\circ$
and is folded to combine the forward and backward intervals.  Monte
Carlo samples are used to obtain the values of the coefficients, which
are chosen to maximize the statistical separation between signal and
background events.  The distributions of $\cal F$ for Monte Carlo
simulated $\Bz\to h^+ h^{\prime -}$ decays and background events in
the \mes\ sideband region $5.20<\mes <5.27\gevcc$ are displayed in
Fig.~\ref{fig:fisher}(a).

The final reconstruction efficiencies range from $31\%$ to
$45\%$, depending on the mode.  The detection efficiencies, which
include the branching fractions of $\Kz\to\KS\to\pip\pim$ and
$\piz\to\gamma\gamma$ \cite{PDG}, are listed in Table \ref{tab:brresults}.

Signal yields are determined from an unbinned maximum likelihood fit
that uses \mes, $\Delta E$, \fish, and
$\theta_c$ (where applicable).  Separate fits are performed for each
of the four topologies, where the likelihood for a given candidate $j$
is obtained by summing the product of event yield $n_i$ and
probability ${\cal P}_i$ over all the possible signal and background
hypotheses $i$.
The $n_i$ are determined by maximizing the extended likelihood
function $\cal L$:
\begin{equation}
{\cal L}= \exp\left(-\sum_{i=1}^M n_i\right)\,
\prod_{j=1}^N \left[\sum_{i=1}^M n_i {\cal P}_i\left(\vec{x}_j;
\vec{\alpha}_i\right)
\right]\, .
\end{equation}
The probabilities ${\cal P}_i(\vec{x}_j;\vec{\alpha}_i)$ are evaluated
as the product of probability density functions (PDFs) for each of the
independent variables $\vec{x}_j$, given the set of parameters
$\vec{\alpha}_i$.  Monte Carlo simulation is used to validate the
assumption that the fit variables are uncorrelated.  The exponential
factor in the likelihood accounts for Poisson fluctuations in the total
number of observed events $N$.  For the $K^\pm\pi^\mp$, $\pi^\pm\piz$,
$K^\pm\piz$, $\KS\pi^\pm$, and $\KS K^\pm$ terms, the yields are
rewritten in terms of the sum $n_{f}+n_{\bar{f}}$ and the asymmetry
$\acp =(n_{\bar{f}}-n_{f})/(n_{\bar{f}}+n_{f})$, where
$n_f\,(n_{\bar{f}})$ is the fitted number of events in the mode $\B\to
f\,({\Bb\to\bar{f}})$.  The numbers of events, $N$, entering the
maximum likelihood fit for each topology are $16032$ ($h^+h^{\prime -}$),
$16452$ ($h^+\piz$), $3623$ ($\KS h^+$), and $1503$ ($\KS\piz$).

The parameters for background \mes\ and $\Delta E$ PDFs are determined from 
events in on-resonance $\Delta E$ sideband regions.  The signal \mes\ and
$\Delta E$ PDF parameters are determined from fully reconstructed
$B^+\to \Dzb \pi^+$ and $B^+\to\Dzb\rho^+$ $(\rho^+\to\pip\piz)$ decays.
Events in on-resonance \mes\ sideband regions and Monte Carlo simulated
signal decays are used to parameterize the Fisher discriminant PDFs
for background and signal, respectively (see Fig.~\ref{fig:fisher}(a)).
Alternative parameterizations obtained from off-resonance data and
Monte Carlo simulation are used as cross-checks and for determination
of systematic uncertainties.
The $\theta_c$ PDFs
are derived from kaon and pion tracks in the momentum range of
interest from approximately $42\,000$ $D^{*+}\to\Dz\pip$ ($\Dz\to
\Km\pip$) decays.  This control sample is used to parameterize the
$\theta_c$ resolution $\sigma_{\theta_c}$ as a function of track
polar angle.  The resulting $K$--$\pi$ separation,
defined as $|\theta_c^K-\theta_c^\pi|/\sigma_{\theta_c}$, where
$\theta_c^K$ $(\theta_c^\pi)$ is the expected Cherenkov angle for
a kaon (pion), is shown as a
function of momentum in Fig.~\ref{fig:fisher}(b).  

The results of the fit are summarized in Table \ref{tab:brresults}, where
the statistical error for each mode corresponds to a $68\%$ confidence 
interval and is given by the change in signal yield $n_i$ that corresponds 
to a $-2\ln{\cal L}$ increase of one unit.  
Signal significance is defined as the square root of the change in
$-2\ln{\cal L}$ with the corresponding signal yield fixed to zero.
For the three modes that have statistical significance less than $4\sigma$
we report Bayesian $90\%$ confidence level upper limits.
In addition, for the purpose of combining with measurements from other
experiments, we report the branching 
fractions corresponding to the fitted signal yields:
$\BR(\btoppz)=(5.1^{+2.0}_{-1.8}\pm 0.8)\times 10^{-6}$,
$\BR(\btoKK)=(0.85^{+0.81}_{-0.66}\pm
0.37)\times 10^{-6}$ and $\BR(\btoKzKp) = (-1.3^{+1.4}_{-1.0}\pm 0.7)\times
10^{-6}$.
The upper limit on the signal yield for mode $i$ is given by the value of
$n_i^0$ for which
$\int_0^{n_i^0} {\cal L}_{\rm max}\,dn_i/\int_0^\infty 
{\cal L}_{\rm max}\,dn_i = 0.90$, where ${\cal L}_{\rm max}$ is 
the likelihood as a function of $n_i$,
maximized with respect to the remaining fit parameters.
Branching fraction upper limits are calculated by increasing the signal
yield upper limit and reducing the efficiency by their respective 
systematic errors.  

\begin{figure}[!tbh]
\begin{center}
\includegraphics[width=1.0\linewidth]{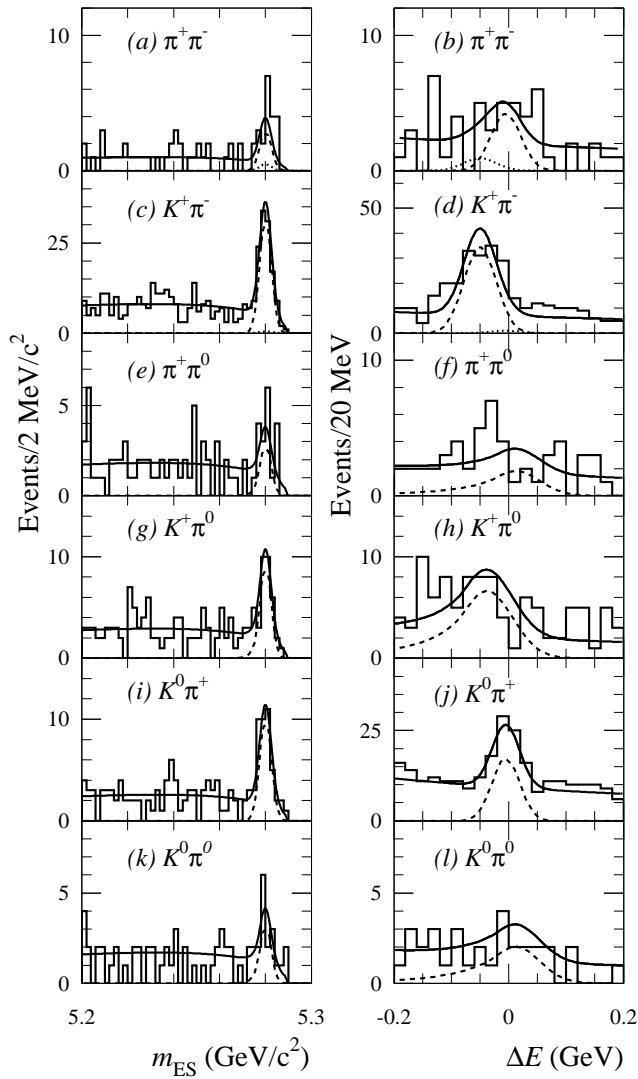}
\caption{
The \mes\ and $\Delta E$ distributions for the various modes, using
likelihood ratio requirements described in the text.
The solid curves represent the fit predictions for both signal and
background; the dashed curve represents the given signal mode only and 
the dotted curve represents other modes of the same topology.
}
\label{fig:prplots}
\end{center}
\end{figure}

Figure~\ref{fig:prplots} shows the distributions in \mes\ and $\Delta E$
for events passing the selection criteria, as well as requirements
on likelihood
ratios, which are used to increase the relative 
fraction of signal events of a given type.  
These likelihood ratios are defined for a given topology
as ${\cal R}_{\rm sig} =
\sum_s n_s {\cal P}_s/\sum_i n_i {\cal P}_i$ and 
${\cal R}_k = n_k{\cal P}_k/\sum_s n_s {\cal P}_s$, 
where $\sum_s$ denotes the sum
over the probabilities for signal hypotheses only, $\sum_i$ denotes
the sum over all the probabilities (signal and background), and
${\cal P}_k$ denotes the probability for signal hypothesis
$k$.  These probabilities
are constructed from all the PDFs except that describing the 
displayed variable.  The likelihood fit projections, scaled by
the relative efficiencies for the likelihood ratio requirements, 
are overlaid on each distribution.  

Systematic uncertainties arise from:
imperfect knowledge of the PDF shapes, uncertainties in the detection
efficiencies, and potential charge bias in track reconstruction and
particle identification.  Uncertainties in the PDF shapes affect both
branching fraction and charge asymmetry measurements.

For most of the branching fraction measurements, the PDF shapes contribute
the largest systematic error.  The exception is the
\btoKpz\ mode, where the largest systematic error is due to the
$5\%$ uncertainty in the \piz\ reconstruction efficiency.
PDF systematic errors are estimated either by varying the PDF
parameters within $1\sigma$ of their measured uncertainties or by
substituting alternative PDFs from independent control samples.
The systematic errors in the signal yields due to PDF 
uncertainties depend on decay mode as shown in 
Table \ref{tab:brresults}.

The $D^{*+}$ control sample of kaon and pion tracks is used to estimate
systematic uncertainties in the asymmetries
arising from possible charge biases in the
$\theta_c$ quality requirements, as well as from differences in 
$\theta_c$ reconstruction for different charge species.  From these
studies we conservatively assign a systematic uncertainty of
$\pm 0.01$ on \acp\ for all the modes.
Charge biases in the detector and track reconstruction chain are studied
in high statistics samples of charged tracks in multihadron events.
These studies show differences in reconstruction efficiencies for
positively and negatively charged tracks of less than $0.005$.  We assign
an overall systematic uncertainty of $\pm 0.01$ on \acp\ for 
possible charge-correlated biases in track reconstruction and particle
identification.
All measured background asymmetries are consistent with zero with statistical 
uncertainties less than $0.03$.  The fitted signal yields and asymmetries
for off-resonance data and on-resonance $\Delta E$ sidebands are also 
consistent with zero.

The overall systematic errors on the branching fractions and charge
asymmetry
measurements are computed by adding in quadrature the PDF systematic
uncertainties 
and the systematic uncertainties on the efficiencies or
due to possible charge biases, respectively.

In summary, we have measured branching fractions for the rare charmless
decays \btopp, \btoKpp, \btoKpz, $\Bu\to\Kz\pip$, and \btoKzpz,
and set upper limits on \btoKK, \btoppz, and \btoKzK.  We find no evidence
for direct \CP\ violation in the observed decays and set $90\%$ C.L.
intervals.  These measurements are in good agreement with previous
results~\cite{cleobr,cleocp}.  

\input{pubboard/acknowledgements}

\vskip12pt
\hbox{$\ \ \ \ \ \ \ \ \ \ \ \ \ \ \ \ \ \ \ \ \ \ \ \ ${\LARGE{\bf---------------}}}

\end{document}

%% file: symbols.tex
\newcommand{\btopp}{\ensuremath{\Bz\to \pip\pim}}

\newcommand{\btoKpp}{\ensuremath{\Bz\to \Kp\pim}}

\newcommand{\btoKK}{\ensuremath{\Bz\to \Kp\Km}}

\newcommand{\btoppz}{\ensuremath{B^+\to \pip\piz}}

\newcommand{\btoKpz}{\ensuremath{B^+\to \Kp\piz}}

\newcommand{\btoKzK}{\ensuremath{B^+\to \Kzb\Kp}}

\newcommand{\btoKzKp}{\ensuremath{B^+\to \Kzb\Kp}}

\newcommand{\btoKzpz}{\ensuremath{\Bz\to \Kz\piz}}

\newcommand{\etal}{{\it et al.}}

\def\fish    {\ensuremath{\cal F}}

\def\acp {\ensuremath{{\cal A}}}

\def\thsph    {\ensuremath{\theta_{\scriptscriptstyle S}}}

%% file: authors_0110.tex
%
\author{B.~Aubert}
\author{D.~Boutigny}
\author{J.-M.~Gaillard}
\author{A.~Hicheur}
\author{Y.~Karyotakis}
\author{J.P.~Lees}
\author{P.~Robbe}
\author{V.~Tisserand}
\affiliation{Laboratoire de Physique des Particules, F-74941 Annecy-le-Vieux, France }
\author{A.~Palano}
\affiliation{Universit\`a di Bari, Dipartimento di Fisica and INFN, I-70126 Bari, Italy }
\author{G.P.~Chen}
\author{J.C.~Chen}
\author{N.D.~Qi}
\author{G.~Rong}
\author{P.~Wang}
\author{Y.S.~Zhu}
\affiliation{Institute of High Energy Physics, Beijing 100039, China }
% ...

%% file: pubboard/acknowledgements.tex
We are grateful for the 
extraordinary contributions of our \pep2\ colleagues in
achieving the excellent luminosity and machine conditions
that have made this work possible.
The collaborating institutions wish to thank 
SLAC for its support and the kind hospitality extended to them. 
This work is supported by the
US Department of Energy
and National Science Foundation, the
Natural Sciences and Engineering Research Council (Canada),
Institute of High Energy Physics (China), the
Commissariat \`a l'Energie Atomique and
Institut National de Physique Nucl\'eaire et de Physique des Particules
(France), the
Bundesministerium f\"ur Bildung und Forschung
(Germany), the
Istituto Nazionale di Fisica Nucleare (Italy),
the Research Council of Norway, the
Ministry of Science and Technology of the Russian Federation, and the
Particle Physics and Astronomy Research Council (United Kingdom). 
Individuals have received support from the Swiss 
National Science Foundation, the A. P. Sloan Foundation, 
the Research Corporation,
and the Alexander von Humboldt Foundation.

%% file: pub_0110_la.bbl
\begin{thebibliography}{99}

\bibitem{ckm} 
N.~Cabbibo, \jprl{10}, 531 (1963);
M.~Kobayashi and T.~Maskawa, Prog. Theor. Phys. {\bf 49}, 652 (1973).

\bibitem{bellestb}
Belle Collaboration, A.~Abashian \etal, \jprl{86}, 2509 (2001).

\bibitem{babarstb}
\babar\ Collaboration, B.~Aubert \etal, \jprl{86}, 2515 (2001).

\bibitem{gamalpha}
M.~Gronau and D.~London, \jprl{65}, 3381 (1990);
M.~Gronau, J.L.~Rosner and D.~London, \jprl{73}, 21 (1994);
R.~Fleischer and T.~Mannel, \jprd{57}, 2752 (1998);
M.~Neubert, J.\ High Energy Phys.\ {\bf 02}, 014 (1999);
M.~Beneke, G.~Buchalla, M.~Neubert and C.T.~Sachrajda, \jprl{83}, 1914 (1999);
M.~Neubert, Nucl.\ Phys.\ Proc.\ Suppl.\ {\bf 99}, 113 (2001).

\bibitem{cc}
Charge conjugate states are assumed throughout, except where explicitly
noted.

\bibitem{cleobr} 
CLEO Collaboration, D.~Cronin-Hennessy \etal, \jprl{85},
515 (2000).

\bibitem{cleocp} 
CLEO Collaboration, S.~Chen \etal, \jprl{85},
525 (2000).

\bibitem{babarnim}
\babar\ Collaboration, B.~Aubert \etal, SLAC-PUB-8569, submitted to
Nucl.\ Instrum.\ and Methods.

\bibitem{pepii}
PEP-II Conceptual Design Report, SLAC-R-418 (1993).

\bibitem{fox}
G.C.~Fox and S.~Wolfram, \jprl{41}, 1581 (1978).

\bibitem{spheric}
S.L.~Wu, Phys.\ Rep.\ {\bf 107}, 59 (1984).

\bibitem{PDG} Particle Data Group, D.E.~Groom \etal, Eur.\ Phys.\ J.\ 
C~{\bf 15}, 1 (2000).

\end{thebibliography}
